\documentclass[twocolumn]{aastex631}
\usepackage{color,soul}
\usepackage{amsmath}
\usepackage{booktabs} 
\usepackage{hyperref}

\begin{document}

\title{Using Neural Networks to Automate the Identification of Brightest Cluster Galaxies in Large Surveys }

\author[0009-0004-9465-428X]{Patrick Janulewicz}

\affiliation{Department of Physics, McGill University, Montréal, Canada}
\affiliation{Trottier Space Institute, McGill University, Montréal, Canada}
\affiliation{Ciela Institute, Montréal, Canada}
\affiliation{Mila - Quebec Artificial Intelligence Institute, Montréal, Canada}

\author[0000-0002-0104-9653]{Tracy M. A. Webb}
\affiliation{Department of Physics, McGill University, Montréal, Canada}
\affiliation{Trottier Space Institute, McGill University, Montréal, Canada}

\author[0000-0003-3544-3939]{Laurence Perreault-Levasseur}
\affiliation{Trottier Space Institute, McGill University, Montréal, Canada}
\affiliation{Ciela Institute, Montréal, Canada}
\affiliation{Mila - Quebec Artificial Intelligence Institute, Montréal, Canada}
\affiliation{Department of Physics, Université de Montréal, Montréal, Canada}
\affiliation{Center for Computational Astrophysics, Flatiron Institute, New York, USA}
\affiliation{Perimeter Institute for Theoretical Physics, Waterloo, Canada}



\begin{abstract}

Brightest cluster galaxies (BCGs) lie deep within the largest gravitationally bound structures in existence. Though some cluster finding techniques identify the position of the BCG and use it as the cluster center, other techniques may not automatically include these coordinates. This can make studying BCGs in such surveys difficult, forcing researchers to either adopt oversimplified algorithms or perform cumbersome visual identification. For large surveys, there is a need for a fast and reliable way of obtaining BCG coordinates. We propose machine learning to accomplish this task and train a neural network to identify positions of candidate BCGs given no more information than multiband photometric images. We use both mock observations from \textsc{The Three Hundred} project and real ones from the Sloan Digital Sky Survey (SDSS), and we quantify the performance. Training on simulations yields a squared correlation coefficient, R$^2$, between predictions and ground truth of R$^2 \approx 0.94$ when testing on simulations, which decreases to R$^2 \approx 0.60$ when testing on real data due to discrepancies between datasets. Limiting the application of this method to real clusters more representative of the training data, such those with a BCG r-band magnitude $r_{\text{BCG}} \leq 16.5$, yields R$^2 \approx 0.99$. The method performs well up to a redshift of at least $z\approx 0.6$.  We find this technique to be a promising method to automate and accelerate the identification of BCGs in large datasets.

\end{abstract}



\section{Introduction} \label{sec:intro}

Galaxy clusters are the largest gravitationally bound objects in the universe. At the heart of these clusters lies the brightest cluster galaxy (BCG). These BCGs are among the most massive and luminous galaxies in the universe today. They also play a pivotal role in our understanding of galaxy evolution. The BCG is often considered to be the optical center of the galaxy cluster, lying at the base of the cluster's gravitational potential well \citep{van_den_Bosch_2005}. 

Some optical cluster finding techniques naturally identify the most likely BCG candidate and set the center of the cluster to be at that point. A well-known example of this is redMaPPer \citep{Rykoff_2014}, which assigns up to 5 candidates ranked by probability. Another example is the method seen in \cite{whl12} (henceforth WHL12), which selects the brightest galaxy within a certain distance from the cluster center.

However, not all cluster finding techniques use the BCG as a center. X-ray surveys \citep{liu_2022,rosat_clusters_vikhlinin}, for example, select the X-ray emission peak instead. Another popular method is to use fluctuations in the cosmic microwave background caused by the Sunyaev-Zeldovich (SZ) effect as tracers of galaxy clusters \citep{Bleem_2015,planck_clusters_2016}. In this case, the center is selected to be the peak of the hot gas distribution. The BCG, however, has been shown to often have an offset with the center found in X-ray or SZ surveys, particularly as redshift increases. These offsets are also influenced by the dynamical state of the cluster, as more disturbed clusters often host BCGs that are less centrally located  \citep{depropris_2021, gozaliasl2019}. Similar results are also suggested by cosmological simulations \citep{Cui2016}. Even in optical surveys, it is not guaranteed that the BCG is automatically identified. For instance, some algorithms select clusters by searching for overdensities of galaxies in redshift or color space. These include the Spitzer Adaptation of the Red-sequence Cluster Survey \citep{Muzzin_2009, sparcs} and the Massive and Distant Clusters of WISE Survey \citep{madcows}. Because these techniques focus on overdensities rather than individual galaxies, the coordinates of the BCG centroid are not necessarily provided.

With recent and upcoming projects such as Euclid \citep{euclid} and the Vera C.\ Rubin Observatory \citep{lsst} expected to greatly increase the quantity of available wide-field data, having a quick and accurate method of identifying the BCG from images will become crucial. Such large amounts of data will make manual identification infeasible, while overly simplistic algorithms may fail given the complexity of the task. As a result, there is a need to better automate this process.

To study a population of brightest cluster galaxies, researchers may opt to construct a catalog containing their positions and properties \citep{chu_2021,kluge_bender_2023,smith_2023}. BCG identification may also be done in simulations \citep{roche_2024}. A few approaches to automate this task have been explored. For instance, work from \cite{Somboonpanyakul_2022} utilizes the probability distribution of redshift and stellar mass for objects near the SZ cluster center to assign BCG likelihoods. On the other hand, methods from \cite{Chu_2022} involve careful treatment of foreground and background objects before selecting the most likely remaining BCG candidate.

In this work, we suggest an automated approach that is capable of appropriately weighing features including optical color, morphology, and size, requiring little more than photometric images. Such requests lend themselves well to the use of machine learning.

We therefore test the abilities of neural networks to identify the BCG centroid from galaxy cluster images. We train a neural network on simulated cluster images from \textsc{The Three Hundred} project \citep{the300} and test its performance on real cluster images from the Sloan Digital Sky Survey (SDSS) \citep{sdss, sdss_dr8}. We also train the neural network on real galaxy cluster images and compare the results. Though we test the network on galaxy clusters up to redshift $z=0.8$ later in the paper, we introduce this work as a proof of concept for a relatively narrow range of low-redshift observations ($z=0.15$ to $z=0.25$). By doing so, we hope to minimize evolutionary effects and limitations caused by resolution or depth, thus reducing systematic uncertainties.

This method is beneficial for several reasons. For one, it works directly on images, circumventing the need to obtain galaxy catalogs, remove foreground and background objects, or perform other preparatory steps. Moreover, it does not require redshift information for each galaxy, but rather an approximate redshift of the cluster. This can be particularly helpful for surveys with limited optical follow-up or high photometric redshift uncertainties. It is also physics-agnostic by nature, which allows it to function without the constraints of predefined models of galaxy behavior. Instead, this method can adapt to any range of cluster environment that is represented in the training dataset.

The option to train on simulations also provides a great deal of advantanges. For instance, simulations allow us to be certain about our choice of BCG. In observations, there is always a chance that the galaxy selected as the BCG is incorrectly identified. This is especially useful for generalizing the task to higher redshift, where the true BCG may be difficult to discern using real observations. In particular, the BCG has been predicted to lose its unique identity around $z\approx 0.7$ \citep{De_Lucia_2007}, as it becomes less certain that the brightest galaxy at this time will become the main progenitor of the redshift zero BCG. Using simulated data also gives access to a variety of characteristics not necessarily known to the observer, allowing for the possibility of further inference. While we recognize that training on mock observations may introduce biases when the simulations stray from reality, we believe that their various benefits make them particularly attractive.

The paper is laid out as follows. In Section \ref{sec:data}, we describe the dataset, which contains both simulated observations and real ones. In Section \ref{sec:methods}, we describe the methods used to process the data, train the machine learning model, and assess its accuracy. In Section \ref{sec:results}, we present the results, which include studying the differences between simulated and real observations, training and testing the model, and performing relevant tests for those wishing to apply these methods to other datasets. In Section \ref{sec:conclusions}, we summarize the contents of the paper and discuss the application of this technique to large surveys.

\section{Data} \label{sec:data}

\begin{figure*}[]
\includegraphics[width=\textwidth]{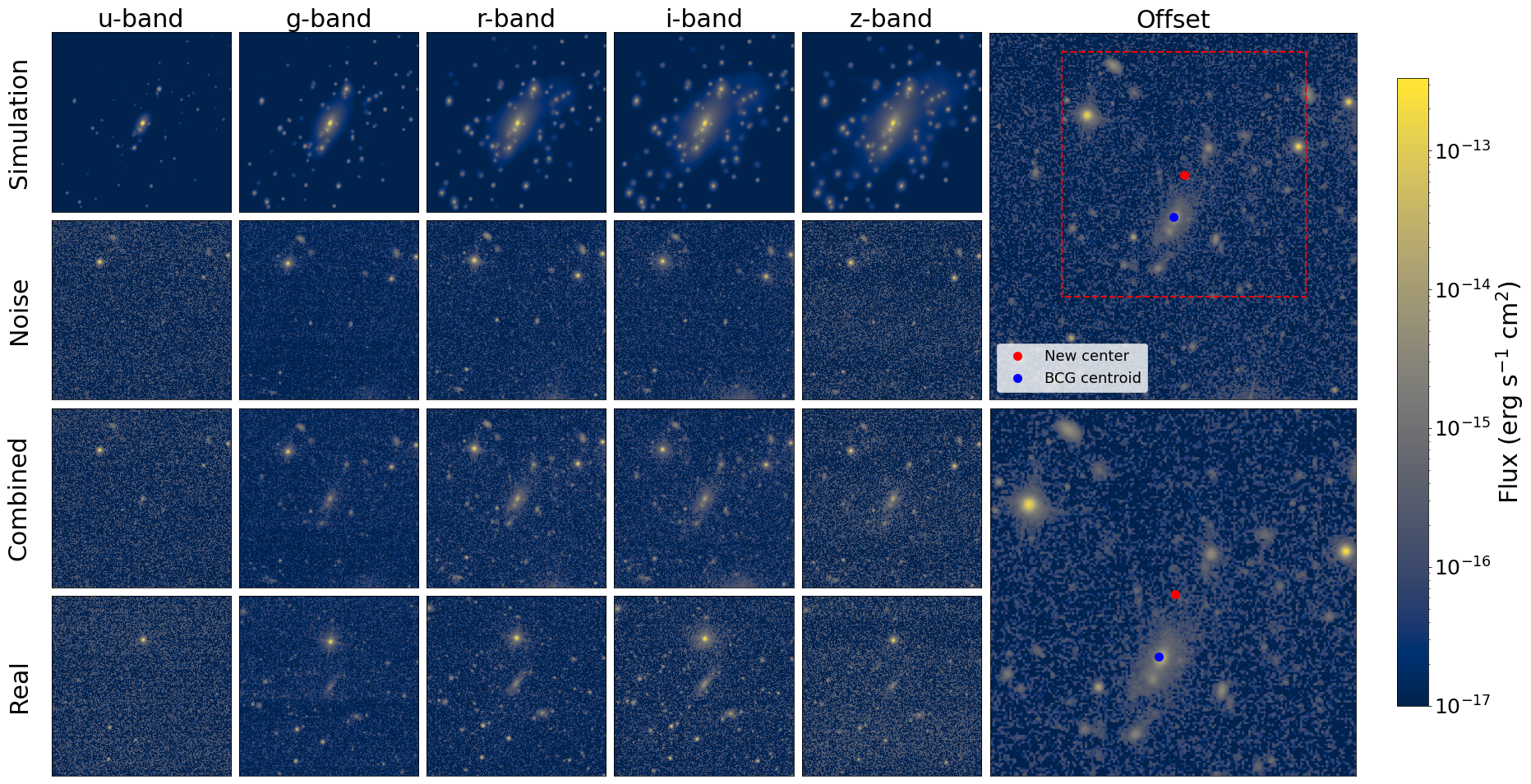}
\caption{The five columns on the left represent photometric bands from SDSS. Simulated galaxy clusters are shown in the first row along with random SDSS noise in the second row. The combination of the two is shown in the third row, and a real WHL12 cluster is shown in the final row for comparison. In the rightmost column, we give an example of a cluster being off-centered by illustrating the process on the simulated r-band.}
\label{fig:data_visual}
\end{figure*}

\subsection{The Three Hundred data \label{subsec:the300_data}}

Simulated clusters are obtained from \textsc{The Three Hundred} project cosmological simulation, which consists of 324 galaxy clusters modeled with different full-physics hydrodynamical re-simulations and semi-analytical models. We use clusters obtained from the \textsc{gizmo-simba} run, as this has been shown to produce more realistic BCGs compared to other codes from the same simulation \citep{gizmo_simba}. We select a redshift range between $z=0.15$ and $z=0.25$. The full dataset of simulated clusters contains four snapshots within this range, all of which are included in our dataset.

Simulated cluster images are then generated using the Python package PyMGal\footnote{\url{https://pypi.org/project/pymgal/} \label{pymgal}}, which calculates magnitudes using techniques from EzGal \citep{ezgal} and then projects them to create mock observations. When using the software, we assume the simple stellar population model described in \cite{bc03} with a Chabrier initial mass function \citep{chab}, and we select the Gaussian smoothing length of a given particle to be the distance to its 30$^{\text{th}}$ nearest neighbor.

For each snapshot, 10 random projection angles are chosen, giving 40 different projections for each of the 324 clusters. This makes for a total of 12960 projections. For each one, we generate images in all five of the SDSS bands. These bands are ultraviolet (u), green, (g), red (r), near-infrared (i), and infrared (z). Each projection is thus made up of a 5-channel 256x256 image. We select the side lengths to represent a physical distance of roughly $1.33$ Mpc. A sample can be found in the first row in Figure \ref{fig:data_visual}.

\subsection{SDSS data \label{subsec:sdss_data}}

The data obtained from SDSS can be split into two separate parts: noise images and cluster images.

We begin with the noise. To make the simulated clusters similar in appearance to observed ones, we must add realism to them. To do this, we superimpose randomly selected images of the sky obtained from SDSS. Coordinates are chosen randomly from the area of the Baryon Oscillation Spectroscopic Survey (BOSS) survey \citep{sdss_boss_dr12}. We then cut out an image with side length equal to approximately $1.33$ Mpc to match the scale of simulated images. This side length is calculated by producing a random redshift between $z=0.15$ and $z=0.25$. For a given redshift, the length is converted to a physical distance by assuming a cosmology of $H_0=67.8$ and $\Omega_{\text{m,0}}=0.307$. The number of unique noise samples collected is equal to the number of simulated observations, meaning that no noise sample is repeated.

Though the mock observations each have a side length of 256 pixels, this is not necessarily true for SDSS data. We must therefore resize SDSS images to match the dimensions of the simulations. We use the rebinning algorithm described in the RealSim Python package \citep{bottrell2019}, which resizes images to a charge-coupled device (CCD) angular scale. This method ensures that the total flux is conserved and maximizes the fidelity of the mock observations. Clean simulated images and SDSS noise are then added together in FITS format as shown in Figure \ref{fig:data_visual}. They are then converted to PNG format using logarithmic scaling.

Images of real galaxy clusters are also obtained from SDSS using the WHL12 catalog. The catalog contains the name, celestial coordinates, and photometric redshift of each galaxy cluster. We use this photometric redshift to select all clusters between $z=0.15$ and $z=0.25$. We then cut out regions with a side length of $1.33$ Mpc and select all images that fit within the boundaries of their SDSS field. We find a total of 6962 real cluster images satisfying these criteria. Images are then processed with the same techniques used for noise images, which include the same rebinning algorithm and pixel dimensions. An example of a real galaxy cluster image is shown at the bottom of Figure \ref{fig:data_visual}.

The WHL12 catalog also contains data describing the BCG brightness and the richness of the cluster. The brightness is defined as the Petrosian magnitude of the BCG in the r-band, which we will refer to as $r_{\text{BCG}}$. The catalog defines a cluster richness $R_{L^*}$ that can be defined mathematically as $R_{L*} = \frac{L_{200}}{L^*}$, where $L_{200}$ is the total r-band luminosity within R$_{200}$ and $L^*$ is the evolved characteristic luminosity of galaxies in the r-band.

Also included is a scaling relation between $R_{L^*}$ and the mass $M_{200}$. This provides us with an approximate relationship between richness and mass, which is shown in the following equation.

\begin{equation}\label{eq:m_vs_rl}
    \log M_{200} = (-1.49 \pm 0.05) + (1.17 \pm 0.03) \log R_{L^*}
\end{equation}

These properties can then be used to isolate subsets of BCGs that align more closely with the simulated dataset. We explore these relationships in greater detail in Section \ref{sec:results}.

\section{Methods \label{sec:methods}}

\subsection{Machine learning details \label{subsec:data_and_model}}

Because both the real and simulated clusters are centered on their BCG, we must modify them to ensure that the model is capable of generalizing to instances where the BCG may be offset, as this will be the case for many surveys. However, we want to instill in the model a tendency to prefer central galaxies, as the BCG in real cluster images should still be near the cluster center defined by the given technique. To do this, we offset the images using a Gaussian perturbation from the center. We randomly draw $x$ and $y$ values from a normal distribution and offset the images by the result. The distribution has a mean of 0 kpc and a standard deviation of 150 kpc. This roughly covers the distribution of offsets found in other works such as the offset between the BCG and the X-ray peak \citep{Seppi_2023}. To ensure that the smaller image does not exceed the boundaries of the larger one, we truncate the Gaussian at the maximum possible offset. We cut out a region of 1 Mpc from the larger $1.33$ Mpc image, meaning that this maximum offset is half the excess space or approximately 167 kpc. The resulting images have pixel dimensions of 192x192.

The images are then divided into training, validation, and test datasets. For real observations, we use 80\% of images for training, 10\% for validation, and 10\% for testing. For the simulations, we split by cluster number to ensure that no simulated galaxy cluster appears in two datasets. These numbers are drawn at random and assigned to a dataset. In other words,  we select 80\% of the 324 clusters for training, 10\% for validation, and 10\% for testing. A breakdown of each dataset can be found in Table \ref{tab:dataset_sizes}.

\begin{deluxetable}{lcccc}[h]
\tablewidth{0pt}
\tablehead{
  \colhead{\textbf{Dataset}} & 
  \colhead{Train } & 
  \colhead{Validation} & 
  \colhead{Test} &
  \colhead{Total}
}
\startdata
\textbf{Primary datasets}\\
\hspace{3mm} Simulation & 10400 & 1280 & 1280 & 12960 \\
\hspace{3mm} Real       & 5569 & 696  & 697  & 6962 \\
\textbf{Comparison datasets} \\
\hspace{3mm} Simulation        & 5000 & -  & -  & - \\
\hspace{3mm} Real    & 5000 & -  & -  & - \\
\hspace{3mm} Real (full redshift)       & 5000 & -  & 5000  & - \\
\enddata
\caption{The number of images across each dataset. Note that the simulated and real comparison datasets are subsets of the corresponding primary dataset. All datasets consist of a redshift range between $z=0.15$ and $z=0.25$ except those in the bottom row specified to cover the full redshift range, which consists of samples between $z=0.05$ and $z=0.8$. \label{tab:dataset_sizes}}
\end{deluxetable}

We then pass the off-centered 5-channel training images to the neural network. We select a ResNet18 architecture \citep{resnet} and pass the $x$ and $y$ pixel values of the BCG centroid as learned parameters. We use a rectified linear unit (ReLU) activation function \citep{relu} and an Adam optimizer \citep{adam}. We select the mean squared error (MSE) as a loss function, which can be expressed mathematically by $\text{MSE} = \frac{1}{n} \sum_{i=1}^{n} \left( \theta_i - \hat{\theta}_i \right)^2$ for a number of samples $n$, a truth vector $\theta$, and a prediction vector $\hat{\theta}$. Further details regarding the model architecture and hyperparameters can be found in Table \ref{tab:model_params}.

\begin{deluxetable}{lc}[ht]
\tablehead{
\colhead{Parameter} & \colhead{Value}
}
\startdata
\textbf{Main model}\\
\hspace{3mm} Architecture & ResNet18 \\
\hspace{3mm} Learning rate & 0.001 \\
\hspace{3mm} Batch size & 16 \\
\hspace{3mm} Optimizer & Adam \\
\hspace{3mm} Activation function & ReLU \\
\hspace{3mm} Loss function & MSE \\
\midrule
\textbf{Transfer learning} & \\
\midrule
\hspace{3mm} Learning rate  & 0.0001 \\
\enddata
\caption{Details and hyperparameters of the model used to determine the BCG position.}
\label{tab:model_params}
\end{deluxetable}

We also use real cluster images to train different networks with the same architecture. We compare one model trained only on simulations, one trained only on real cluster images, and one trained on both via transfer learning. For the transfer learning model, we partially train on simulated images and use the result to initialize the network's weights. We then decrease the learning rate by one order of magnitude and fine-tune on real images. This allows the model to gain information on both simulated and real observations. More details on these tests can be found in Section \ref{sec:results}.

\subsection{Assessing accuracy \label{subsec:assessing_acc}}

We now define the statistics used to quantify the model's accuracy. The first statistic is the coefficient of determination, denoted R$^2$. To assess the quality of the predictions, we can plot the values predicted by the model as a function of the ground truth. A strong correlation indicates that predictions rarely deviate from the truth, meaning that a perfectly accurate model would produce R$^2=1$. A mathematical description of these statistics is shown below, where we denote predicted values with a hat operator, averages with an overline, and true values without an operator symbol. We take $n$ to be the number of images in the relevant test dataset.

\begin{align}
    \text{R}^2_x &= 1 - \frac{\sum_{i=1}^{n} (x_i - \hat{x}_i)^2}{\sum_{i=1}^{n} (x_i - \bar{x})^2} \\
    \text{R}^2_y &= 1 - \frac{\sum_{i=1}^{n} (y_i - \hat{y}_i)^2}{\sum_{i=1}^{n} (y_i - \bar{y})^2}
\end{align}

Another way of quantifying success is by measuring the proportion of predictions that are correct within some reasonable error. In this work, we consider a prediction to be correct if it lies within some distance of the true position and incorrect otherwise. Such a threshold should be large enough to encompass the entire BCG, but small enough to avoid false positives. Unless otherwise specified, we will use a threshold of 25 kpc, as this is a conservative estimate for the region of the cluster dominated by the BCG  \citep{brough}. We define the proportion of predictions that are correct within the distance threshold and refer to it as the accuracy A$_{\text{T}}$. This accuracy can be more formally defined by the equation that follows.

\begin{align}
    \text{A}_{\text{T}} &= \frac{1}{n} \sum_{i=1}^{n} \textbf{1}_{\text{$\leq$ T}}(d_i)
\end{align}

Note that $d_i$ is equal to the Euclidean distance $\sqrt{(x_i - \hat{x}_i)^2 + (y_i - \hat{y}_i)^2}$ between the true and predicted positions. We use $\textbf{1}_{\text{$\leq$ T}}(d_i)$ to denote the indicator function, which is equal to one if the error is less than the threshold and zero otherwise.


\section{Results \label{sec:results}}

\subsection{Quantifying differences between datasets}\label{subsec:dataset_diffs}

Before proceeding to train and test the model, it is crucial to understand potential differences between the real and simulated clusters. To begin, we consider two tests to summarize these differences, which can be done directly on FITS images before they are offset and converted to PNG. To perform these tests, we randomly select 5000 images from each dataset, thus ensuring a large sample size that is consistent across the two groups.

\begin{figure*}[ht]
\centering
\includegraphics[width=\textwidth]{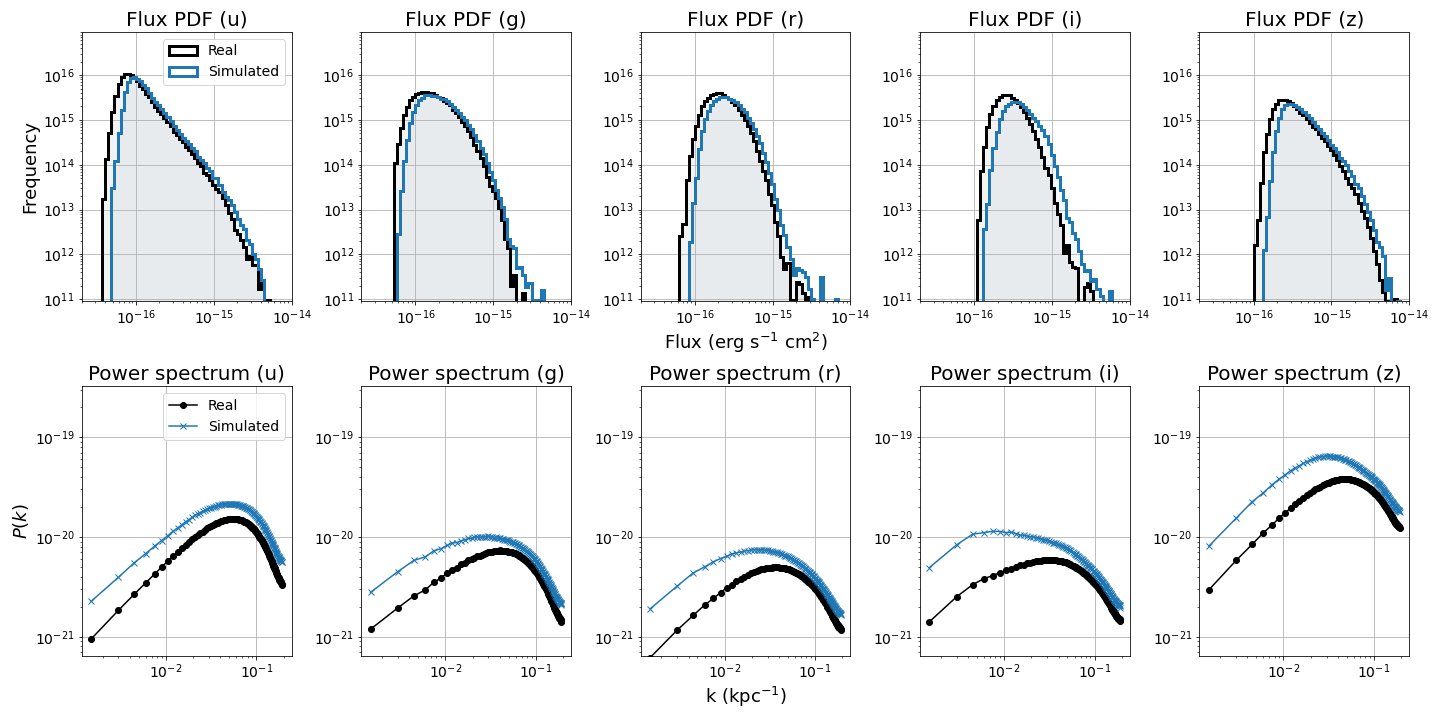}
\caption{The first row contains the 1D PDF of flux values averaged over 5000 images for each dataset. The second row shows the power spectrum averaged over the same images for each dataset. Columns indicate SDSS bands. While the two datasets appear to be quite similar, the simulated images show higher flux and power. \label{fig:summary_stats}}
\end{figure*}

The first test is to study the one-dimensional probability density function (1D PDF) of the flux values. This PDF can be obtained by binning the pixels of each image in a given band to a histogram and then averaging the resulting histograms over all images. Because this is repeated over a large number of experiments, the result should demonstrate the typical brightness distribution of a galaxy cluster image. We obtain a PDF in each band for both simulated and real images, and we show the result in the first row of Figure \ref{fig:summary_stats}.

The second test is to compute the power spectrum of the images. The power spectrum indicates the typical variability between two pixel values at a given length scale. The power, which we denote $P(k)$, quantifies the amount of variability between samples for a given wavenumber $k$. This wavenumber is a spatial frequency representing how often a wave pattern repeats per unit distance. In this case, the largest wavenumber represents the smallest distance, which corresponds to one pixel or approximately 5 kpc. The smallest wavenumber represents the largest distance, which corresponds to half the image dimension or approximately 667 kpc. For each of the 5000 images, we compute a power spectrum in each of the five bands. We then average the resulting power spectra across all images and report the result. The results are shown in the second row of Figure \ref{fig:summary_stats}. 

The flux PDFs demonstrate a tendency for the simulated cluster images to be slightly brighter than the real ones. This is repeated across all SDSS bands. This may indicate that the population of simulated clusters is more luminous than the population of real ones.

The power spectra show that the simulated cluster images show slightly higher variability overall. This is particularly true for smaller wavenumbers, meaning that large distances vary more in simulated images than in real ones. This may also suggest that the simulations show brighter centers, as the larger distances compare BCGs with the outskirts of the galaxy cluster. We will take these differences into consideration in Section \ref{subsec:testing} by isolating subsets of clusters with more prominent BCGs.

\begin{figure}[ht]
\centering
\includegraphics[width=\columnwidth]{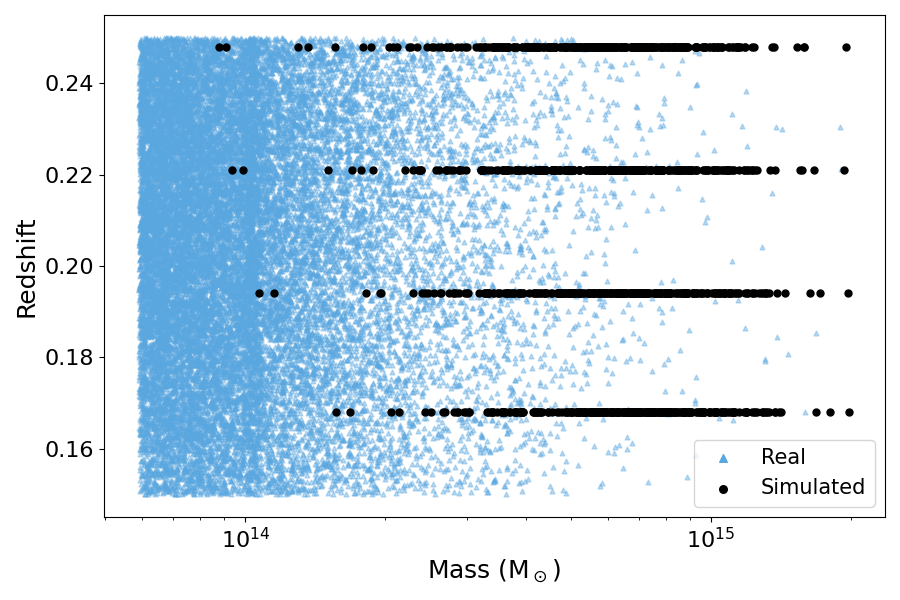}
\caption{The redshift and mass distribution of galaxy clusters. Real clusters are represented by blue triangles, while simulated ones are represented by black circles. Note that the horizontal lines from simulations are due to the four discrete snapshots. Overall, the real clusters cover a wider mass range and tend to be less massive as a whole compared to simulated ones.  \label{fig:m_vs_z}}
\end{figure}

Another key difference is the way in which mass and redshift are distributed between the two datasets. Because simulations are captured via discrete snapshots in time, we expect the real clusters to be far more varied in their redshift. Furthermore, differences in mass cuts can result in significant discrepancies in cluster mass. To visualize these details, we plot the redshift and mass for each. We combine the training, validation, and test datasets for both the real and simulated galaxy clusters to obtain the entire population, and we show the results in Figure \ref{fig:m_vs_z}.

As expected, the real clusters are found to occupy a wide range of values, while the space covered by simulated clusters tends to be more limited. Redshifts from simulations are grouped into thin horizontal lines, while those from real clusters are more evenly distributed. We also see that the simulated clusters tend to be considerably more massive than those found in WHL12. While many low-mass clusters are present in the real data, they are mostly absent from the simulations.

\subsection{Training the model \label{subsec:training}}

We study the model's behavior during training via a learning curve. The learning curve tracks the loss as a function of training epochs. In machine learning, the loss refers to the penalty for incorrect predictions and can be used to indicate the model's accuracy. Training epochs refer to the number of complete passes through the dataset, during which the network's weights are adjusted using gradient descent. The learning curve shows how quickly and how well the model learns to make correct predictions.

\begin{figure}[ht]
\centering
\includegraphics[width=\columnwidth]{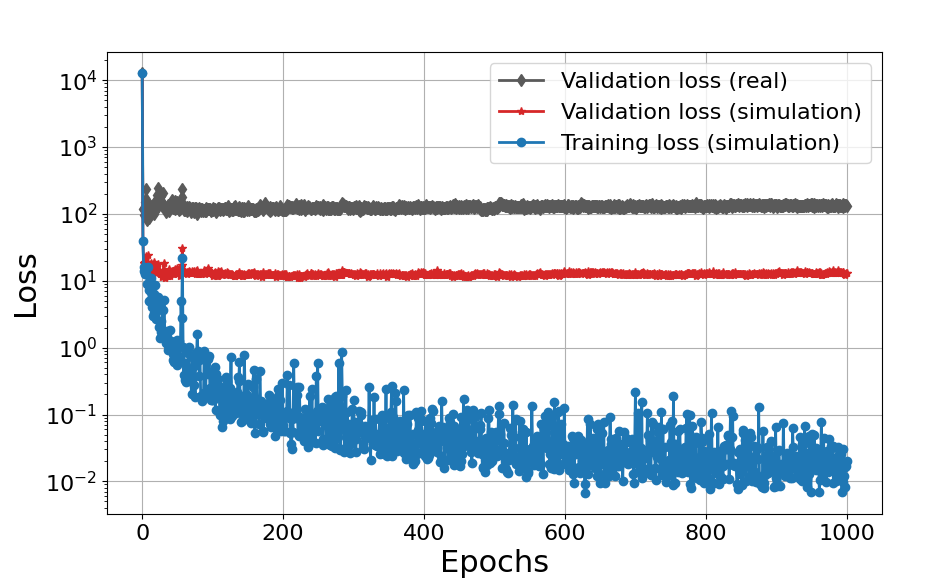}
\caption{Learning curve when training on simulated galaxy
clusters. We compare the training and validation loss for simulated clusters. We also add the validation loss for real clusters. Both validation curves plateau after relatively few epochs, with the validation loss for real cluster images remaining well above that of simulated images. \label{fig:learning_curve}}
\end{figure}

Examining this curve can provide important insights about the model's training process. For instance, learning curves help identify when the model begins to overfit the training set, resulting in difficulty generalizing to unseen data. Preventing overfitting is particularly important for this task, as the simulations contain a limited number of unique training samples.

The learning curve shown in Figure \ref{fig:learning_curve} tracks the loss in three different datasets. The training and validation set of simulated galaxy cluster images make up two of these datasets. The third is the validation set of real galaxy clusters. We can therefore look at both the real and simulated validation sets to identify potential overfitting.

We draw two notable conclusions from this test. The first conclusion is that performance on both validation sets appears to plateau after relatively few epochs. We find that training beyond this point does not significantly reduce the validation loss. To avoid overfitting, we will use a model trained for 100 epochs for testing unless otherwise specified. The second conclusion concerns the relationship between the curves for the two validation sets. Despite both curves plateauing, the loss for the real clusters stays consistently above the loss for the simulated ones. It appears that additional training is insufficient to properly reduce the loss on real images.

\subsection{Testing the model \label{subsec:testing}}

\begin{figure}[ht]
\centering
\includegraphics[width=\columnwidth]{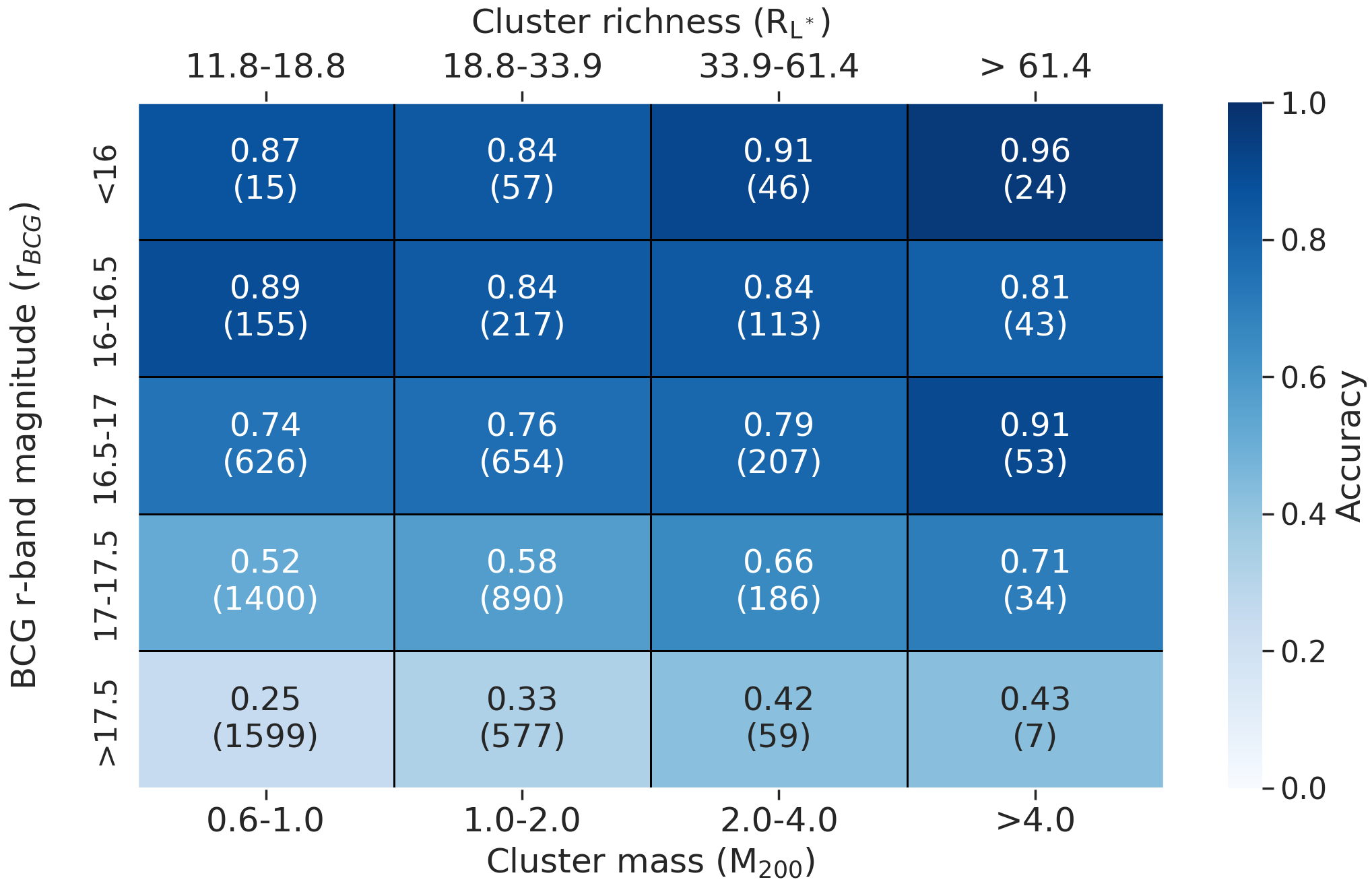}
\caption{Performance for different values of $r_{\text{BCG}}$ and $M_{\text{200}}$. Accuracy is defined as the proportion of predictions within 25 kpc of the true position. The number of samples in each bin is indicated in parentheses. Lower boundaries on ranges are inclusive, while upper boundaries are exclusive. The $R_{L^*}$ richness values corresponding to the $M_{200}$ mass values are shown at the top. In most cases, an increasing trend in accuracy can be seen with greater brightness and greater mass. \label{fig:heatmap}}
\end{figure}

We now test the simulation-trained model on real cluster images. To account for the differences between the two datasets, we can divide the SDSS clusters into groups based on their physical properties. We divide according to the BCG r-band magnitude $r_{\text{BCG}}$ and the cluster mass $M_{200}$, which can be inferred from the richness $R_{L^*}$ using the scaling relation shown in Equation (\ref{eq:m_vs_rl}). Grouping clusters by BCG brightness will allow us to separate prominent BCGs from subtler ones. Meanwhile, separating clusters by mass will allow us to isolate ranges that better align with the simulations used in training. We determine the fraction of predictions accurate within 25 kpc for each subset and show the results in Figure \ref{fig:heatmap}. While all other experiments in this subsection are performed on the test dataset only, we use all 6962 real images in Figure \ref{fig:heatmap}, as we require a large sample to cover all subsets.

While predictions on bright and massive clusters are strong, accuracy drops steeply when testing on clusters with lesser $r_{\text{BCG}}$ and $M_{\text{200}}$. These results raise a few interesting points. The first is that it is possible to identify subsets of the data where accuracy will be strong. When testing on these subsets, we can have confidence that the model will be reliable. When venturing beyond this subset, we must be more cautious. The second conclusion, which aligns with the findings from Section \ref{subsec:dataset_diffs}, is that the simulations do not appear to encompass all possible galaxy clusters. Rather, there appears to a lack of dim BCGs in the simulated training data. This is not entirely unexpected, as we only have 324 unique galaxy clusters. Nonetheless, this limits our ability to generalize well to the wide range of SDSS observations.

Upon inspection of Figure \ref{fig:heatmap}, it appears that the brightness of the BCG has a larger impact on accuracy than cluster mass. While predictions do improve as cluster mass increases, particularly for clusters with dim BCGs, $M_{200}$ appears to be a less reliable indicator of accuracy than $r_{\text{BCG}}$. As a result, we consider a subset of the brightest BCGs available in the WHL12 catalog, which we henceforth define as having $r_{\text{BCG}} \leq 16.5$. The test dataset contains 60 clusters satisfying this requirement.

\begin{figure}[]
\includegraphics[width=\columnwidth]{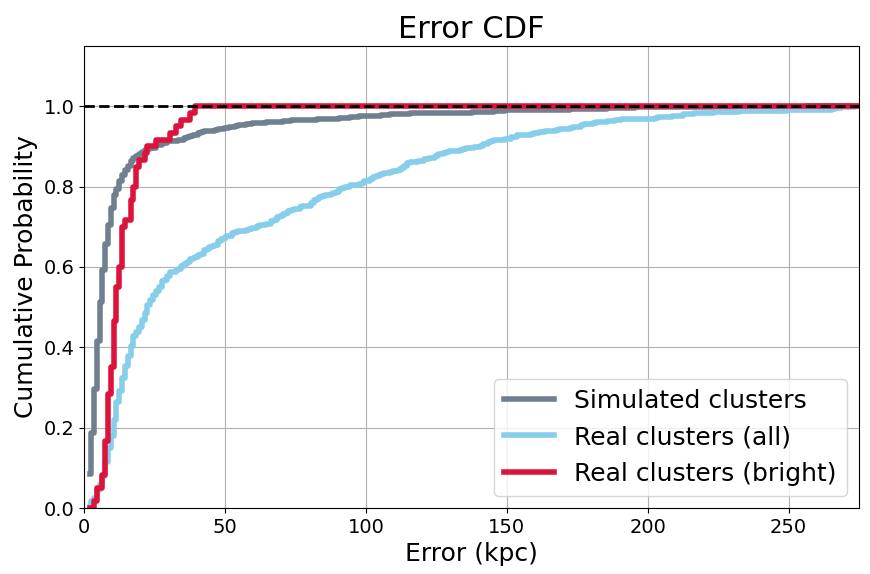}
\includegraphics[width=\columnwidth]{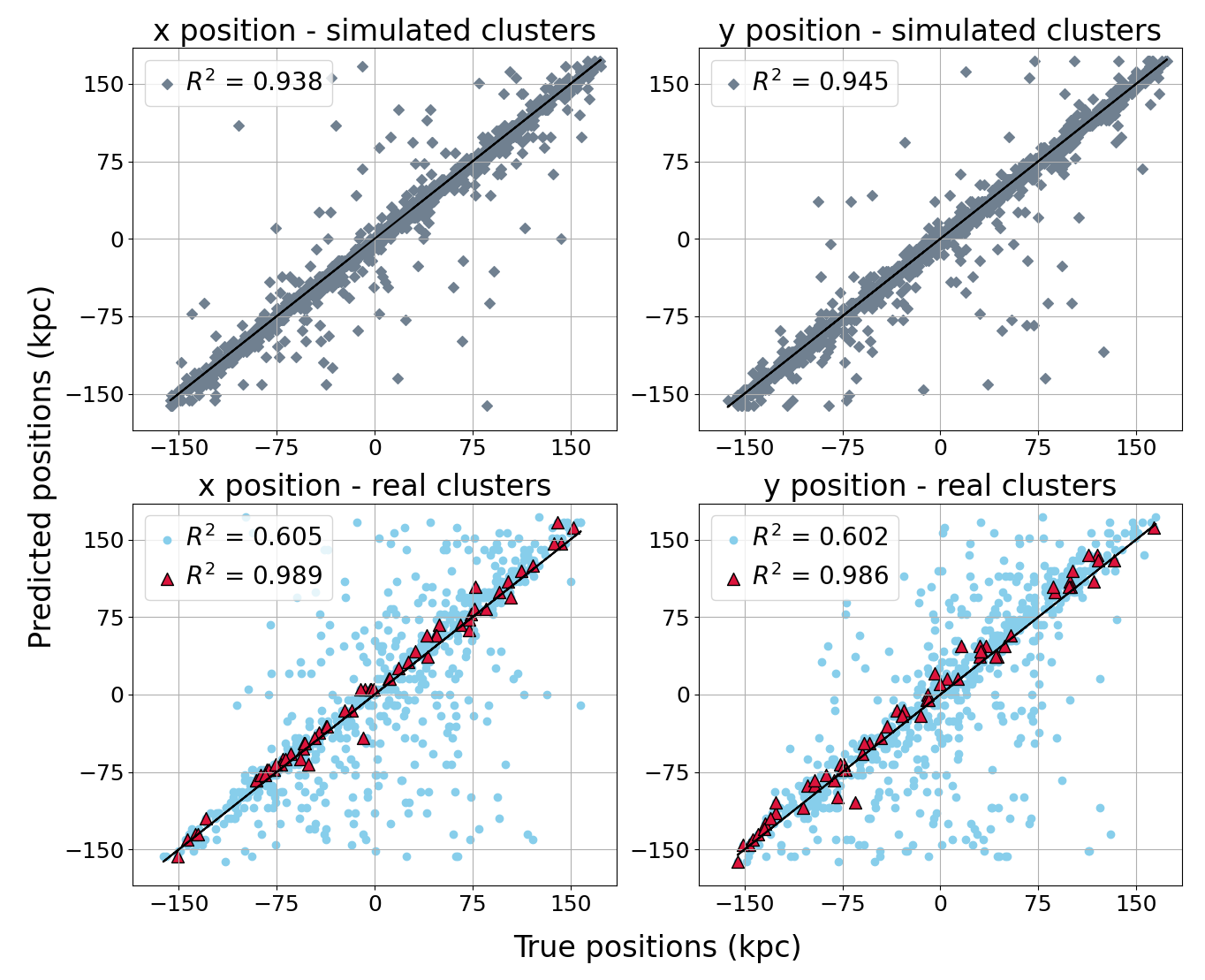}
\caption{Performance when trained on simulated clusters. The model is tested on simulated clusters, real SDSS clusters, and a bright subset of these SDSS clusters satisfying $r_{\text{BCG}} \leq 16.5$. Displayed on top is a cumulative distribution function (CDF) of the Euclidean error across the three test sets. The four figures on the bottom show the true BCG positions plotted against the model's predictions, with a black line displaying a 1:1 relationship shown for reference. The left and right columns show the results for x and y, respectively. Performance is strong when tested on simulations and the bright subset, but it drops when tested on the full set of real SDSS clusters. \label{fig:true_vs_pred}}
\end{figure}

To further assess performance, we study how error is distributed among the model's predictions. We compute the Euclidean distance from the true BCG centroid to the value predicted by the model. We obtain the predicted positions in pixel space and convert them to physical positions. We can then examine the way this error is spread using a cumulative distribution function (CDF). This function plots the fraction of predictions with an error less than or equal to a given value. A CDF that increases steeply near the origin indicates that most predictions are fairly accurate. On the other hand, a function that increases slowly shows that more predictions are inaccurate. The error CDF can be found at the top of Figure \ref{fig:true_vs_pred}.

The CDF shows a discrepancy between the model's predictions across the two datasets. The simulated images tend to have very low error. In fact, we see that most predictions are accurate within roughly 25 kpc. When we look at the real images, this proportion drops by nearly half. This suggests that the model is succeeding in properly identifying some BCGs, but failing in other cases.

We also test performance by plotting the values predicted by the model as a function of the ground truth. The resulting correlation strength will then serve as a measurement of success. We use R$^2_x$ and R$^2_y$ as defined in Section \ref{subsec:assessing_acc} to quantify this strength.

Plots of the true and predicted results can be found at the bottom of Figure \ref{fig:true_vs_pred}. The model trained on simulated images performs well when tested on simulated images, though we find a far weaker correlation on real ones. While many of the predictions fall near the true values, we see far more outliers. We test on the subset of real clusters with $r_{\text{BCG}} \leq 16.5$ and find the correlation strength to be comparable and even superior to tests on simulations. 

In both cases, a trend can be seen when training on simulations and testing on real data. Predictions are highly accurate for bright BCGs, but they become less reliable for dimmer ones. Though much of this error is due to the neural network, it is worth noting that dim BCGs should be more challenging to identify for the WHL12 algorithm as well. As a result, looking at subtler cases can make the ground truth less reliable, therefore adding an additional source of uncertainty.

Moving on from this test, we also look into correcting these biases with transfer learning. Ideally, we would like the model to perform well without having ever seen real data. This would make it possible to pre-train a model for upcoming surveys and apply it as soon as there is a new data release. If this is not possible, we would at least hope to be able to fine-tune the model with small amounts of data.

\begin{figure}[ht]
\includegraphics[width=\columnwidth]{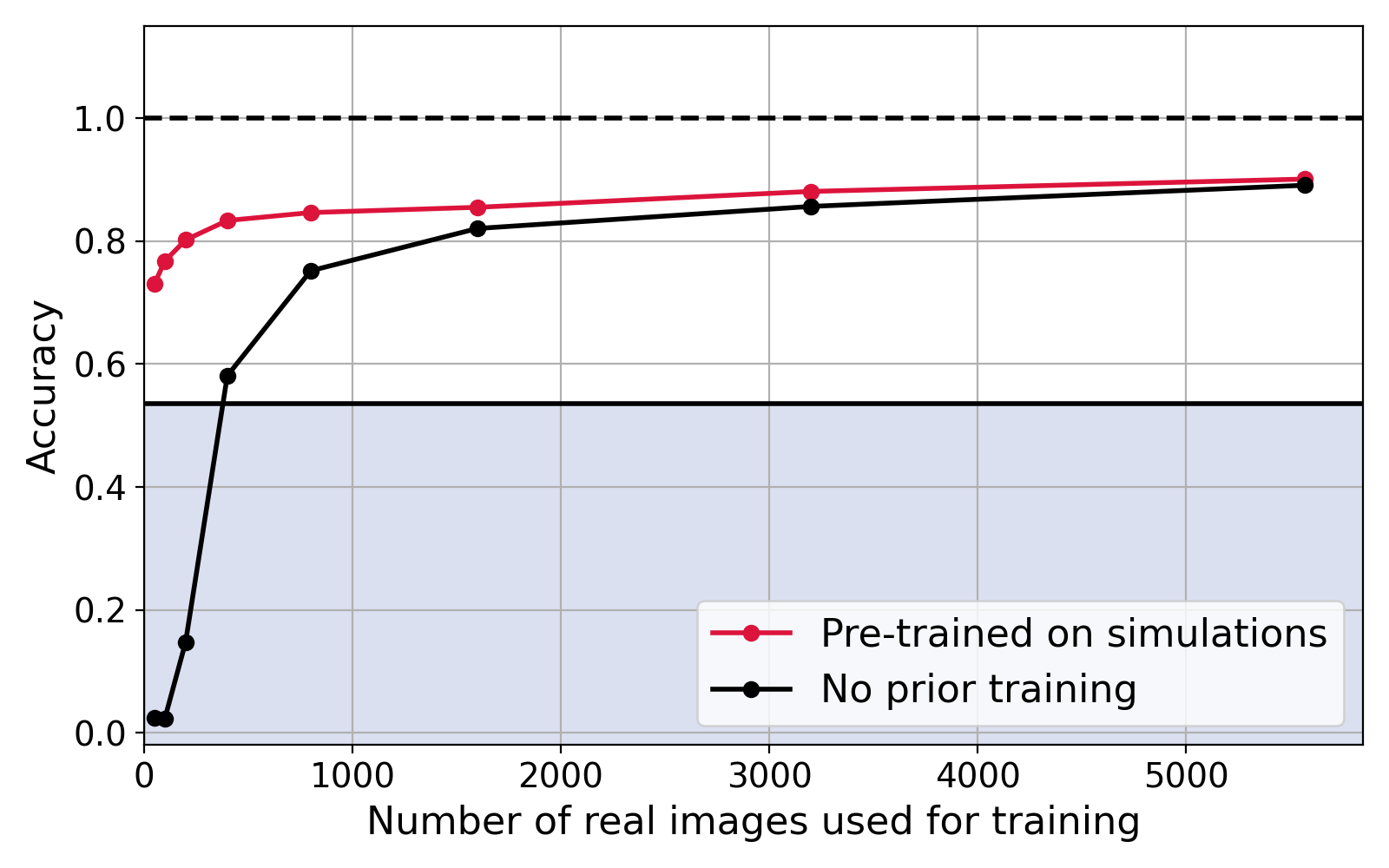}
\caption{The accuracy within a threshold of 25 kpc as a function of training set size. We compare a model trained from scratch on real images to a model pre-trained on simulated images and then fine-tuned on real ones. The solid black line indicates the accuracy for the pre-trained model before transfer learning is applied, with worse values falling in the shaded region below it. Though both models perform well with only a few hundred images, training from scratch requires a larger set of training samples. \label{fig:transfer_learning}}
\end{figure}

We test the model's accuracy with different numbers of real images used for training. We begin with a small set of 50 randomly selected samples. We then continue to double this number until it reaches a size greater than half of the training dataset, in which case we use all available real training images. In each case, we fine-tune for 100 epochs with a learning rate reduced by an order of magnitude as indicated in Table \ref{tab:model_params}. The fine-tuned transfer learning model can then be compared with a model trained from scratch. Results are shown in Figure \ref{fig:transfer_learning}.

This fine-tuned model achieves a strong performance with very few training samples, suggesting that biases in the simulations can be corrected with small amounts of data. In the event that the simulations for large surveys are biased compared to real observations, researchers may be able to correct for this with limited early release data. It is also noteworthy that the gap in performance closes significantly after only a few hundred training images. While simulations have better ground truth and more available information, real data may potentially be less biased. We find that either method can be viable once a modest amount of data is obtained.

\subsection{Comparing different dataset configurations}

We also analyze the model's ability to adapt to different configurations of training and test datasets. One configuration is to train and test on real galaxy cluster images. Another is to invert the problem by training on real images and testing on simulations. If the model trained on real data performs badly when testing on simulations, and the model trained on simulations performs badly when testing on real data, then the issue is likely due to bias between the two datasets. However, if the former model performs well, and the latter model performs poorly, this may indicate that the simulated clusters form a subset of the real ones. Finally, we are also interested in whether we can improve performance using transfer learning as described in Section \ref{subsec:data_and_model}.

\begin{figure}[ht]
\includegraphics[width=\columnwidth]{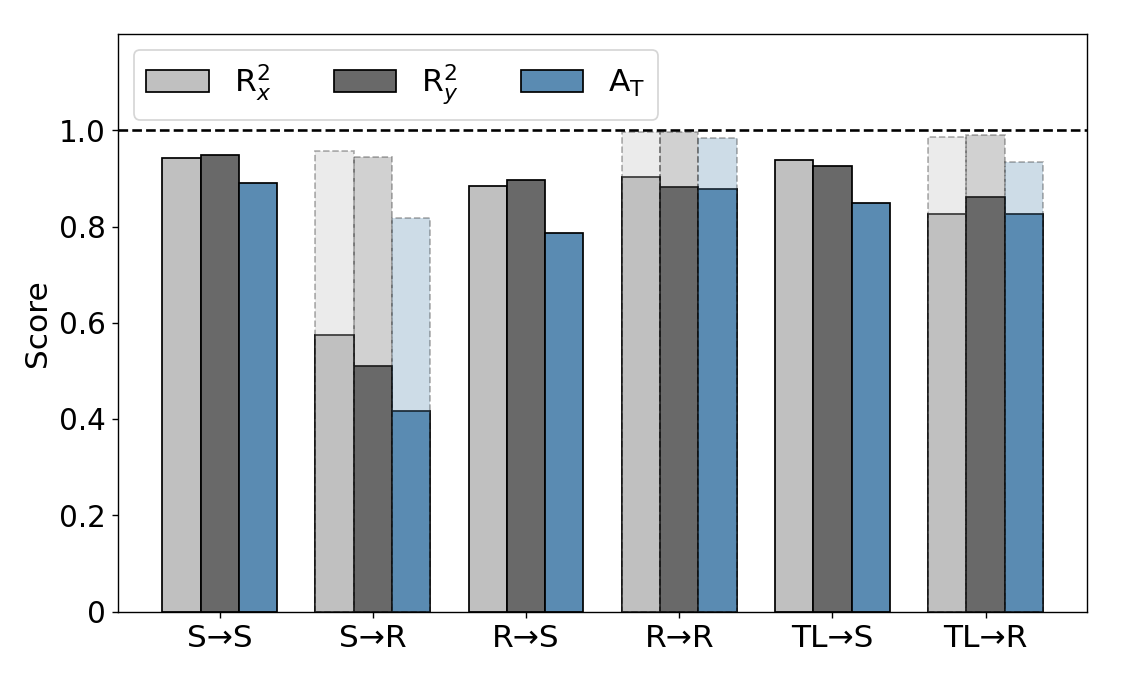}
\caption{A comparison of R$^2_x$, R$^2_y$, and A$_\text{T}$ for different networks tested on the real and simulated cluster datasets. The threshold distance for A$_\text{T}$ is set to be 25 kpc. Labels to the left of the arrow represent the dataset used for training. They are abbreviated S for simulated images, R for real images, and TL for transfer learning. Labels to the right of the arrow represent the test dataset and are abbreviated similarly. Performance on the bright subset of BCGs ($r_{\text{BCG}} \leq 16.5$) is indicated by shaded regions for comparison. Results are strong in all cases except for the model trained on simulations and tested on real images, although they improve drastically when tested on the bright subset. \label{fig:model_comp}}
\end{figure}

We train three new models on the simulated and real comparison datasets having redshift $0.15 \leq z \leq 0.25$ as indicated in Table \ref{tab:dataset_sizes}. We train one network on 5000 simulated cluster images for 100 epochs and a second network on 5000 real cluster images for 100 epochs. For transfer learning, we take the model trained on 5000 simulated images for 100 epochs and fine-tune it on 1000 real images for another 100 epochs. We show the results in Figure \ref{fig:model_comp} by comparing the three statistics defined in Section \ref{subsec:assessing_acc}.

The model trained on real clusters works well when tested on real clusters, and the model trained on simulations performs well when tested on simulations. Transfer learning shows strong performance across both datasets. When testing on a different dataset than the training set, we get mixed results. We continue to see a drop in performance when training on simulations and testing on real data. However, the model trained on real images appears to perform well when tested on simulations, further suggesting that the simulations form a subset of the real images. We once again isolate the subset of real observations satisfying $r_{\text{BCG}} \leq 16.5$. All networks, regardless of training dataset, make high-accuracy predictions on this subset.

\subsection{Comparing to alternative approaches}

We now discuss the advantages of this approach compared to alternative methods of automatically identifying BCGs. By requiring only photometric images, this approach reduces the need for data products that may be more challenging to obtain, such as accurate redshift values. In the absence of available and accurate galaxy redshifts, separating foreground objects from cluster members becomes more challenging. One straightforward approach in this scenario is to find the brightest galaxy in the cluster's field of view and select this as the BCG.

However, the presence of foreground interlopers can make this a questionable assumption. We therefore investigate how often the brightest galaxy in the field of view is the true BCG by running tests on simulations. For the neural network, we take the simulation-trained model used in Section \ref{subsec:testing} and test it on the simulated cluster images, which have been off-centered and converted from FITS to PNG. For the alternative approach, we take the same simulated test dataset with the same offsets, but we leave the images in FITS format. We then run SExtractor \citep{sextractor} on the r-band using default parameters with a detection threshold of $1.5\sigma$ above the background noise level. To isolate galaxies and remove contamination from stars, we remove objects that do not satisfy $\text{CLASS\_STAR}  \leq 0.1$. We then select the brightest galaxy in the r-band based on the MAG\_AUTO parameter.

\begin{figure}[ht]
\includegraphics[width=\columnwidth]{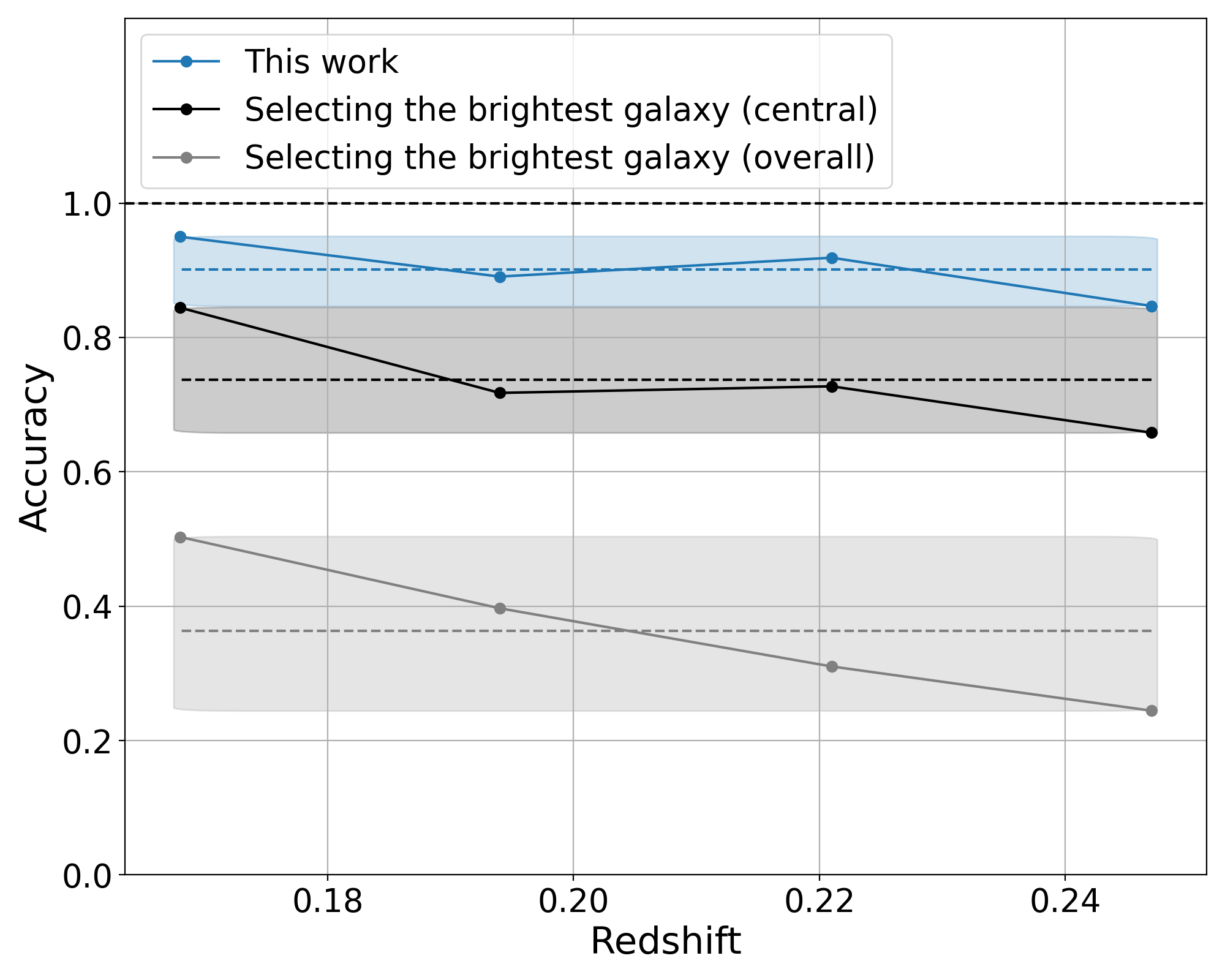}
\caption{The blue curve represents the accuracy of the neural network used in this work for each simulation snapshot. The black curve represents the accuracy achieved by selecting the brightest galaxy in the r-band within 200 kpc of the cluster center, while the gray curve represents the accuracy achieved by selecting the brightest galaxy in the entire field of view. Shaded regions indicate the spread between the highest and lowest accuracy, while the dotted lines within them represent the average accuracy over all redshifts. All tests are performed on off-centered simulated images with a side length of 1 Mpc, with estimates considered to be correct if the predicted position is within 25 kpc of the true value. The neural network achieves the highest accuracy and sees the smallest decrease in performance at higher redshift. \label{fig:cnn_vs_brightest}}
\end{figure}

We consider two similar tests. The first is to select the brightest galaxy in the entire field of view, which has a side length of 1 Mpc. The other approach is to select the brightest galaxy near the center of the cluster. Because the distribution of offsets is known, we can select an appropriate radius of 200 kpc from the center of the image. However, it is worth noting that this may be a more hazardous design choice in real data, where the BCG may have a wider range of possible offsets. The results of this test can be found in Figure \ref{fig:cnn_vs_brightest}.

The neural network outperforms selecting the brightest galaxy in the field of view, highlighting its ability to recognize the difference between BCGs and bright foreground galaxies. Even when imposing an optimal search radius based on a known distribution of offsets, the neural network achieves the highest accuracy for every snapshot. Another notable detail is that the neural network shows the smallest decrease in accuracy when redshift is increased. This is likely due to the fact that foreground contamination becomes a greater challenge at higher redshift. In all cases, the neural network outperforms selecting the brightest galaxy and appears to be better suited to adapt to more challenging cases.

\subsection{Pushing toward higher redshift}

\begin{figure}[ht]
\includegraphics[width=\columnwidth]{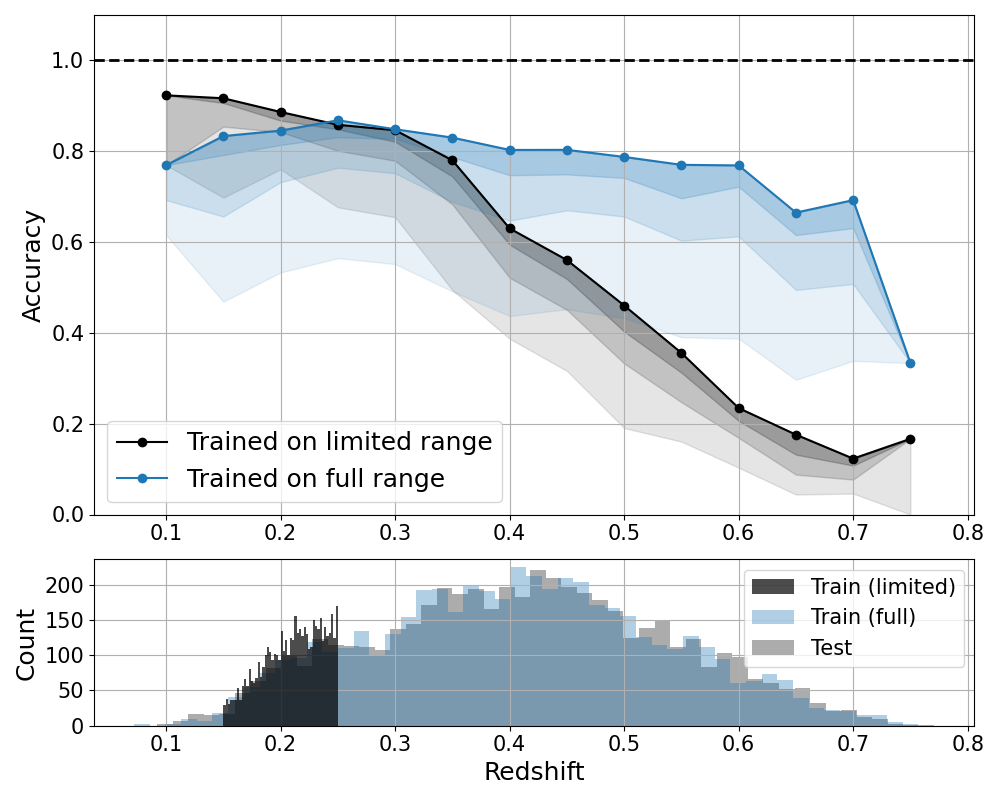}
\caption{The proportion of predictions within a given distance to the ground truth at various redshifts. The model trained on a range of $0.15 \leq z \leq 0.25$ is compared to a model trained over the entire redshift range of the catalog. Both are tested on real SDSS clusters. Solid lines represent the accuracy computed with a threshold of 25 kpc. Shaded regions below the solid lines indicate a tighter error threshold. Ordered from darkest to lightest, the breaks between each shaded region indicate a threshold of 20 kpc, 15 kpc, and 10 kpc. A histogram showing the redshift distribution of each dataset is shown at the bottom. Both models tend to perform well within the training range, though the task becomes increasingly difficult at higher redshift. \label{fig:redshift}}
\end{figure}

Finally, we test the model's ability to adapt to higher redshift and identify the point at which it breaks down. We test two different models, both trained on the real comparison datasets shown in Table \ref{tab:dataset_sizes} to avoid bias from the simulations. The first model is trained on 5000 real galaxy clusters between $z=0.15$ and $z=0.25$. The second one is trained on the same number of cluster images for the same number of epochs, but covering the entire redshift range. We then create a test dataset by collecting 5000 test images over the full redshift range that do not overlap with either training set. In other words, there is an equal number of images used in training and testing for this particular experiment. All images are processed using the same methods described in Section \ref{sec:data}, including the same rebinning algorithm, pixel dimensions, and offset distribution. Results can be seen in Figure \ref{fig:redshift}.

The neural network trained on the limited redshift range shows a gradual decrease as we move toward higher redshift. Results begin to deteriorate immediately as we push beyond the training values, and they decline steeply around $z=0.35$. Even within the training range, there appears to be slightly better performance for closer objects. By the tail end of the redshifts available in the catalog, accuracy has decreased to nearly zero.

We compare this to the network trained on the entire range. This model  shows much better adaptability, as performance does not sharply decrease until roughly $z=0.6$. It does not perform as well on low-redshift clusters as the first, suggesting that there is some benefit to a more specialized model. This also shows that the task can be performed at somewhat higher redshift without significant decreases in accuracy.

Neither model is capable of accurately predicting the BCG position at the tail end of the range. While it is true that these BCGs may be more subtle than those nearby, this is not necessarily the reason for the drop in accuracy. Limitations in depth and resolution from SDSS likely play a role as well. To properly study this phenomenon, observations from more modern telescopes may be required.

\section{Discussion \label{sec:conclusions}}

Neural networks show promising results when tasked with identifying brightest cluster galaxies in multiband photometric images. Models trained on simulated images can identify the simulated BCGs with near-perfect accuracy, achieving a correlation strength of R$^2 \approx 0.94$ when plotting the predicted values as a function of the truth. Perhaps more impressive is that training and testing on real galaxy cluster images shows similar success. While the simulations are limited to 324 unique galaxy clusters and their BCGs, each telescope observation obtained is truly unique. We therefore expect a greater challenge, as there is greater variability in the clusters and the galaxies that form them. Real data also introduces the possibility of false detections or other uncertainties, which could further hurt performance. Despite these challenges, we obtain accurate predictions even when training on as few as 1000 samples. 

The biggest challenge arises when attempting to bridge the gap between datasets due to differences such as mass cuts and the typical brightness of BCGs. This makes it difficult for the model to generalize between the two datasets even when precautions are taken to avoid overfitting. The quality of predictions from the simulation-trained model drops when tested on real clusters, yielding a correlation strength of R$^2\approx 0.60$. Nonetheless, we find some degree of success when training and testing on different datasets. The model trained on real observations produces strong results when tested on simulations. The converse, however, is not true, suggesting that the simulations cover only a subset of reality.

We are able to isolate a subset of more obvious cases where the network trained on simulations shows extremely strong results. A simple way of doing this is by taking only clusters with particularly bright BCGs, such as those with a BCG r-band magnitude $r_{\text{BCG}} \leq 16.5$. In this case, the correlation strength increases to R$^2 \approx 0.99$. The accuracy within 25 kpc is also high in this subset, with the percentage of successes ranging from the eighties to even the high nineties in some cases. These tests show that it is possible for a neural network trained on simulations to identify BCGs in real observations, suggesting that this task can also be done in anticipation of upcoming surveys. Though preparing for these surveys by training on simulations is not trivial, it can be done with great success given the correct dataset. Those wishing to use such an algorithm should take these findings into consideration and thoroughly test the model's adaptability to unseen data. One possible strategy to address this challenge is to assign a confidence level or an uncertainty to a given prediction. Software such as Python's simulation-based inference (SBI) package \citep{tejero-cantero2020sbi} exists for these purposes. In surveys where the ground truth is not necessarily known, estimating uncertainties may help isolate simple cases from more challenging ones.

Even in the event of slight differences between datasets, transfer learning can be employed to fine-tune the network and drastically improve performance. Biases in the model can easily be corrected with only a few hundred samples.

Given a trained model, BCGs can be accurately identified with simple photometric images and without the need for significant pre-processing, such as handling foreground and background objects. This approach strongly outperforms selecting the brightest galaxy in the field of view, as it is capable of discerning between bright foreground interlopers and the true BCG. The option to train on simulations also opens interesting possibilities for identifying high-redshift BCGs or BCG progenitors.

We also see a level of adaptability to higher redshift. While results are best when testing on the same redshift range used in training, the boundaries can be pushed to some degree without major drops in accuracy. Training over a wider range is also a viable option, but there is a limit around $z\approx 0.6$ at which performance drops steeply. Given the depth and resolution of SDSS, it is unclear whether this limit is the result of technological limitations or the evolution of the galaxy clusters. To settle this, a similar experiment would need to be performed on a deeper, higher-resolution survey.

We reiterate that this method shows strong results, and that neural networks are highly effective for identifying brightest cluster galaxies in large surveys. The primary challenge facing this method is adapting between simulated and real observations. We encourage individuals performing this task to exercise caution when training on simulations, as the quality and diversity of these simulations have a significant impact on the model's adaptability to real observations. While bridging the gap is indeed possible, it is challenging due to the limited amount of unique data from cosmological simulations. Those applying this task to upcoming surveys should consider these facts before moving forward.






\section*{Acknowledgments}


We acknowledge support from the Centre de recherche en astrophysique du Québec, un regroupement stratégique du FRQNT. TW and PJ acknowledge the support of the Natural Sciences and Engineering Research Council of Canada (NSERC), [funding reference number RGPIN-2020-04606]. LPL acknowledges support from the Canada Research Chair Program. We also acknowledge access to the theoretically modeled galaxy cluster data via \textsc{The Three Hundred} collaboration \footnote{\url{https://www.the300-project.org}}. The simulations used in this paper have been performed in the MareNostrum Supercomputer at the Barcelona Supercomputing Center, thanks to CPU time granted by the Red Española de Supercomputación. As part of \textsc{The Three Hundred} project, this work has received financial support from the European Union’s Horizon 2020 Research and Innovation program under the Marie Skłodowska-Curie grant agreement number 734374, the LACEGAL project.



%




\bibliography{sample631}{}
\bibliographystyle{aasjournal}



\end{document}